\documentclass[review]{elsarticle}
\usepackage{amssymb}
\usepackage[utf8]{inputenc}
\usepackage{natbib}
\usepackage{graphicx}
\usepackage{caption}
\usepackage{subcaption}
\usepackage{tabularx}
\usepackage{color}
\usepackage{cite}
\usepackage{lineno}
\usepackage{hyperref}
\usepackage{multirow}
\usepackage[final]{pdfpages}
\usepackage{fancyvrb}
\usepackage{amsmath,amsfonts,amsthm,amssymb}
\usepackage{setspace}
\usepackage{Tabbing}
\usepackage{color}
\usepackage{fancyhdr}
\usepackage{lastpage}
\usepackage{extramarks}
\usepackage{chngpage}
\usepackage{soul,color}
\usepackage{graphicx,float,wrapfig}
\usepackage{framed}
\usepackage{color}
\usepackage{epstopdf}
\usepackage{capt-of}
\usepackage{mathrsfs} 
\usepackage{graphicx}
\usepackage{braket}
\setcitestyle{square}
\journal{Physica A}

\begin{document}

\begin{frontmatter}


 \author{Chinmoy Samanta}
 \ead{samantachinmoy111@gmail.com}

\title{Reaction-Diffusion dynamics in presence of  active barrier: Pinhole sink}

\address{School of Basic Sciences, Indian Institute of Technology Mandi,
Kamand, Himachal Pradesh, 175005, India }

\begin{abstract}
\noindent In this article, we give a semi-analytic expression for survival probability when particles are diffusing in an active potential well. There is no analytic solution available in the literature, due to the requirement of inverse Laplace transform of the propagator, when a sink is placed at the uphill of the parabolic potential even in case of the localized sink. We also explain some of the physical aspects by using our solution. 
\end{abstract}


\end{frontmatter}


\section{INTRODUCTION}
For a long time, the study of diffusion of a particle in the potential well facing active barrier has been an interesting aspect in reaction-diffusion theory \citep{burada2012escape,bagchi2012molecular,hanggi1990reaction,kramers1940brownian,skinner1978relaxation,grote1980stable}. Even the one dimensional model itself is employed to various chemical and biological phenomena involving chemical reaction in condensed phase, electron \citep{marcus1985electron}, and proton \citep{samanta1990deuterium} transfer processes. In semiconductor physics, electron and hole are facing active barrier for the conduction of current \citep{tamura2013potential}, tunneling in semiconductor \citep{chang1974resonant}. From the theoretical point of view, the problem is to calculate the probability that the particle will still be in the excited state, after a time t. The major obstacle of developing the theory for similar kind of phenomena was the complex form of the propagator of the relevant equation, and that hinder the finding of the Laplace transform even in consideration of localized initial distribution as well as localized sink. In ref. \citep{samanta1993exact}, they provided the equation in the Laplace domain for calculating the rate constant. However, the exact expression for survival probability is still not unveiled. Although, recently, a real-time numerical result of the probability is provided by Spendier $et$ $al.$ \citep{spendier2013reaction}. In Fig 4, they showed that the character of the survival probability is deviating drastically for the uphill and downhill location of the sink with respect to the initial particle location. The fascinating nonmonotonic effect has been noticed when particle faces active barrier (uphill placement of sink) with varying steepness of the parabolic potential.
\newline In this article we provide a semi-analytic expression for the survival probability $Q(t)$ of the particle undergoing in parabolic potential and the localized sink is placed at the uphill relative to centrally placed initial distribution. In order to get the expression in the time domain, we have considered an approximation of the Smoluchowski time domain propagator but remain the same in the long-time limit. So our study is limited by beyond the initial time. An interesting nonmonotonic phenomenon is observed as in the ref. \citep{spendier2013reaction}. The exact and approximate form of the propagator is compared graphically in the Appendix.  

\section{SMOLUCHOWSKI EQUATION AND SURVIVAL PROBABILITY}
We begin with a generalization of the diffusion equation, viz., the Smoluchowski equation,
\begin{equation}
    \frac{\partial P(x,t)}{\partial t} = \frac{\partial}{\partial x}(\gamma x P(x,t)+D \frac{\partial P(x,t)}{\partial x})-S(x)P(x,t).
\end{equation}
In the above $P(x,t)$ describe the probability density of finding the particle, executing random walk, at position $x$ and time $t$. $D$ stands for diffusion coefficient, causes diffusive motion. $S(x)$ is position dependent sink term on the potential energy curve (PEC). The rate of motion towards the centre of attractive potential is defined by $\gamma$ and the motion can be called as potential induced motion. The propagator of the above equation in the absence of sink term is nothing but Green's function of the equation can be found by the method of characteristics[] with considering the initial feeding is localized, $P(x,0)=\delta(x-x_{0})$, which has the form 
\begin{equation}
    G(x,t|x_{0})=\frac{Exp[-(x-x_{0} Exp[-\gamma t])^{2}/4D\tau(t)]}{\sqrt{4 \pi D \tau(t)}},
\end{equation}
where $\tau$ is given by
\begin{equation}
    \tau(t)=\frac{1-Exp[-2 \gamma t]}{2 \gamma}.
\end{equation}
The survival probability is 
\begin{equation}
    Q(t)=\int_{-\infty}^{+\infty}P(x,t)dx.
\end{equation}
 The Laplace transform of Eq. (1), which is  
\begin{equation}
     [s-\mathscr{L}+S(x)]\mathscr{P}(x,s)=P(x,0),
\end{equation}
where
\begin{equation}
      \mathscr{L}=(D\frac{\partial^2}{\partial x^2}+\frac{\partial}{\partial x}\gamma x).
\end{equation}
Here $\mathscr{P}(x,s)$ is the Laplace transform of $P(x,t)$. The solution in terms of Green's function of the Eq. (5) is given by
\begin{equation}
     \mathscr{G}(x,s|x_{0})=<x| [s-\mathscr{L}+S(x)+]^{-1}|x_{0}>.
     \end{equation}
By using the operator identity of quantum mechanics we can get
     \begin{equation}
         [s-\mathscr{L}+S(x)]^{-1}=[s-\mathscr{L}]^{-1}-[s-\mathscr{L}]^{-1}S(x)[s-\mathscr{L}+S(x)]^{-1}.
         \end{equation}
For a localized capture of the particle the sink function is chosen to be Dirac delta function, $S(x)= k_{1}\delta(x-x_{1})$. $k_{1}$ is being the strength of the sink function. Thus,       
\begin{equation}
\mathscr{G}(x,s|x_{0})=\mathscr{G}_{0}(x,s|x_{0})-\frac{k_{1}\mathscr{G}_{0}(x,s|x_{1})\mathscr{G}_{0}(x_{1},s|x_{0})}{1+k_{1}\mathscr{G}_{0}(x_{1},s|x_{1})}.
\end{equation}
In the above $\mathscr{G}_{0}(x,s|x_{0})$, the Green's function corresponding to no sink on the PEC and it is nothing but Laplace transform of $G(x,t|x_{0})$, is defined as\\
\begin{equation}
    \mathscr{G}_{0}(x,s|x_{0})=<x|[s-\mathscr{L}]^{-1}|x_{0}>.
\end{equation}
The survival probability in the Laplace-domain would be
\begin{equation}
    \tilde{Q}(s)=\frac{1}{s}[1-\frac{\mathscr{G}_{0}(x_{1},s|x_{0})}{\frac{1}{k_{1}}+\mathscr{G}_{0}(x_{1},s|x_{1})}].
\end{equation}
Our calculation start with combining Eq. (11) and Eq. (2) In the next section.

\section{NON CENTRALLY PLACED TRAP: APPROX. ANALYTIC SOLUTION}
Due to the complex form of the Eq. (2), it is difficult to get the analytical solution in the time domain even for the localized sink of arbitrary position. Although methods are available to get the analytic solution in Laplace domain for the generalized sink (by considering collection Dirac delta function) on PEC. We consider the case when the initial distribution is localized at the center of the potential(consider to be the origin without loss of generality), and position of the localized sink is anywhere except the center. Therefore Eq. (11) become
\begin{equation}
    \tilde{Q}(s)=\frac{1}{s}[1-\frac{\mathscr{G}_{0}(x_{1},s|0)}{\frac{1}{k_{1}}+\mathscr{G}_{0}(x_{1},s|x_{1})}].
\end{equation}
We need to find the Laplace transform of $G(x,t|x_{0})$ by putting $x_{0}=0$ in order to get $\mathscr{G}_{0}(x_{1},s|0)$. We have the expression 
\begin{equation}
    G(x,t|0)= \sqrt{\gamma/(2 \pi D)} (1-Exp[-2 \gamma t])^{-1/2} Exp[-(x^{2}\gamma/(2 D))/(1-Exp[-2 \gamma t])].
\end{equation}
In case of the limit beyond initial time, we can write

\begin{equation}
    G(x_{1},t|0)=  \sqrt{\frac{\gamma}{2 \pi D }}e^{-\frac{x_{1}^2 \gamma}{2 D}}(1-Exp[-2 \gamma t])^{-1/2} Exp[-\frac{x_{1}^2 \gamma}{2 D}\times Exp[-2 \gamma t]].
\end{equation}

One can find the Laplace transform of Smoluchowski propagator, the above equation, from a table of Laplace transform \citep{bateman1954tables,roberts1966table} or otherwise, that
\begin{equation}
    \mathscr{G}_{0}(x_{1},s|0)= \frac{e^{-\frac{3 x_{1}^{2} \gamma}{4 D}}}{\sqrt{8 \pi D \gamma}} \frac{\Gamma(\frac{1}{2}) \Gamma (\frac{s}{2\gamma})}{\Gamma(\frac{1}{2}+ \frac{s}{2 \gamma})}(-\frac{x_{1}^{2}\gamma}{2 D})^{-\frac{1}{4}(1+\frac{s}{\gamma})}M_{1/4-s/4\gamma,s/4 \gamma-1/4}(-\frac{x_{1}^{2}\gamma}{2 D}).
\end{equation}
In the above $\Gamma(n)$ is the gamma function and the $M$ in Eq. (16) is the Whittaker $M$ function define in ref. \citep{abramowitz1965handbook} as 
\begin{equation}
    M_{\kappa,\mu}(z)=e^{-z/2}z^{\mu+1/2}M(1/2+\mu-\kappa,1+2\mu,z),
\end{equation}
where
\begin{equation}
    M(a,b,z)=\frac{\Gamma(b)}{\Gamma(b-a)\Gamma(a)}\int_{0}^{1}dt e^{z t}t^{a-1}(1-t)^{b-a-1}.
\end{equation}
For the case of other propagator, $\mathscr{G}_{0}(x_{1},s|x_{1})$, of Eq. (12) under the approximation, it becomes 
\begin{equation}
\mathscr{G}_{0}(x_{1},s|x_{1})=\frac{e^{-\frac{x^2 \gamma}{2 D}}}{ \sqrt{8 \pi D \gamma}} B(\frac{s}{2 \gamma},\frac{1}{2}),
\end{equation}
where $B(z,w)=\Gamma(z)\Gamma(w)/\Gamma(z+w)$ is the beta function. We consider the case of perfect absorption for that $k_{1}\rightarrow{}\infty$. Substitution of Eq. (20) and Eq. (16) into Eq. (12) gives an expression for survival probability in Laplace domain, which is
\begin{equation}
    \tilde{Q}(s)=\frac{1}{s}-\frac{e^{-\frac{x^2 \gamma}{4 D}}\frac{\Gamma(\frac{1}{2}) \Gamma (\frac{s}{2\gamma})}{\Gamma(\frac{1}{2}+ \frac{s}{2 \gamma})}(-\frac{x_{1}^{2}\gamma}{2 D})^{-\frac{1}{4}(1+\frac{s}{\gamma})}M_{1/4-s/4\gamma,s/4 \gamma-1/4}(-\frac{x_{1}^{2}\gamma}{2 D})}{s B(\frac{s}{2 \gamma},\frac{1}{2})}.
\end{equation}
\begin{figure}[hbt!]
    \centering
    \includegraphics[width=8cm]{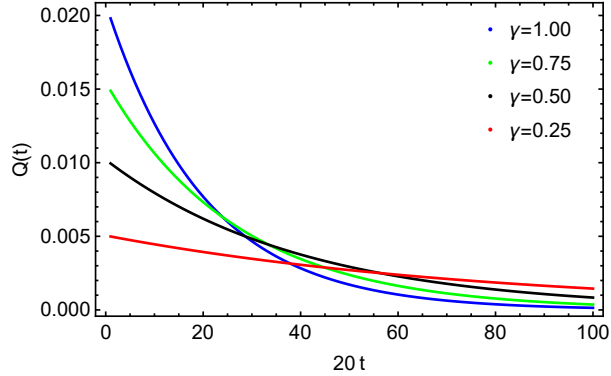}
   \caption{\small{\emph{Survival probability $Q(t)$ Vs $t$ with $x1 = 2;
D = 100$. We see the interesting feature when the placement of initial distribution is at the center of the potential and the trap is elsewhere. We get the nonmonotonic effect as the decay of $Q(t)$ is decreased by increasing the potential steepness but then enhanced on further increase in time. }}}
\end{figure}
The denominator of the second term can be written as
\begin{equation}
    \frac{1}{s B(\frac{s}{2 \gamma},\frac{1}{2})} = \frac{1}{2 \gamma ( \Gamma(1/2))^2}B(\frac{s}{2 \gamma}+\frac{1}{2},\frac{1}{2}).
\end{equation}
Farther we can write the Eq. (21) as
\begin{equation}
     \tilde{Q}(s)=\frac{1}{s}-\tilde{f}_{1}(s)\tilde{f}_{2}(s)
\end{equation}
where
\begin{equation}
\begin{split}
     \tilde{f}_{1}(s)&=\frac{\Gamma(\frac{1}{2}) \Gamma (\frac{s}{2\gamma})}{\Gamma(\frac{1}{2}+ \frac{s}{2 \gamma})}(-\frac{x_{1}^{2}\gamma}{2 D})^{-\frac{1}{4}(1+\frac{s}{\gamma})}e^{-\frac{x^2 \gamma}{4 D}}M_{1/4-s/4\gamma,s/4 \gamma-1/4}(-\frac{x_{1}^{2}\gamma}{2 D});\\
     \tilde{f}_{2}(s)& = \frac{B(\frac{s}{2 \gamma}+\frac{1}{2},\frac{1}{2})}{2 \gamma(\Gamma(1/2))^2}.
\end{split}
\end{equation}
After doing inverse Laplace transform from the table we get the general semi analytic expression in convolution form for $Q(t)$ is given by
\begin{equation}
    Q(t)=1- \int_{0}^{t}f_{1}(t_{0})f_{2}(t-t_{0})dt_{0}.
\end{equation}

Here $f_{1}(t)$ and $f_{2}(t)$ are inverse Laplace transform of $\tilde{f}_{1}(s)$ and $\tilde{f}_{2}(s)$ respectively. It is easy to find the inverse Laplace transform of $\tilde{f}_{1}(s)$ and we use a property of inverse Laplace transform to find $f_{2}(t)$ from $\tilde{f}_{2}(s)$ \citep{roberts1966table}. The property is 
\begin{equation}
    L^{-1}[\tilde{g}(as-b)]=\frac{1}{a}g(\frac{t}{a})e^{\frac{b}{a}t},
\end{equation}
where a, b are constant. The explicit form of $f_{1}(t)$ and $f_{2}(t)$ are
\begin{equation}
    \begin{split}
        f_{1}(t) & =\frac{2 \gamma}{\sqrt{1-Exp[-2 \gamma t]}}Exp[-\frac{x_{1}^{2}\gamma}{2 D}Exp[-2 \gamma t]];\\
         f_{2}(t) & =\frac{Exp[- \gamma t]}{(\Gamma(1/2))^2\sqrt{1-Exp[-2 \gamma t]}}
    \end{split}
\end{equation}
\begin{figure}[hbt!]
    \centering
    \includegraphics[width=8cm]{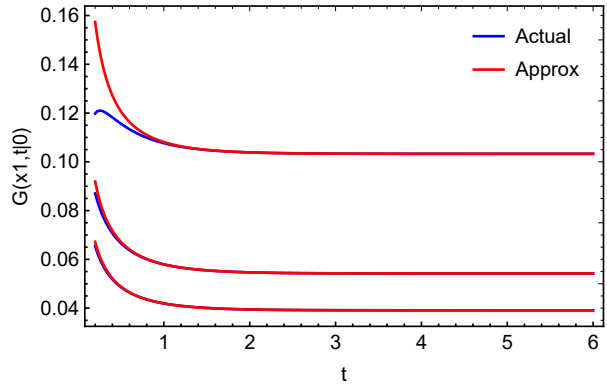}
    \caption{\small{\emph{Actual (blue) and approx.(red) value of $G(x_{1},t|x_{0})$  Vs $t$ for different value of $D$. From the above to the bottom, the curves correspond to $D=10$, $50$ and $100$ respectively. Other fixed parameters are $\gamma = 1$, $x_{1} = 2$ and $x_{0} = 0$.}}}
\end{figure}
\begin{figure}[hbt!]
    \centering
    \includegraphics[width=8cm]{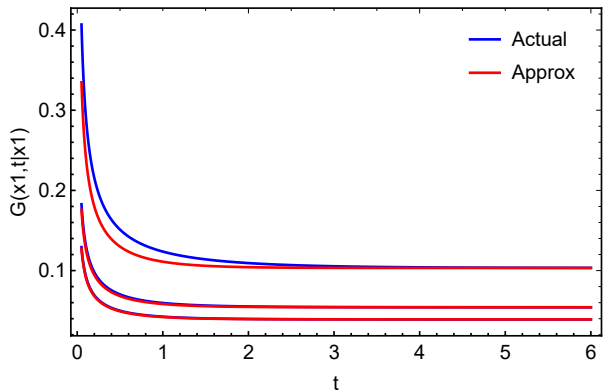}
   \caption{\small{\emph{Actual (blue) and approx.(red) value of $G(x_{1},t|x_{1})$  Vs $t$ for different value of $D$. From the above to the bottom, the curves correspond to $D=10$, $50$ and $100$ respectively. Other fixed parameters are $\gamma = 1$ and $x_{1} = 2$.}}}
\end{figure}
The characteristics of survival probability with the different strength of the potential are demonstrated in Fig. 1. As it is prohibited to consider the short time limit of our analysis, we interpret in the high time limit. It is merely depicted that the behavior of $Q(t)$ is as expected as in the ref. \citep{spendier2013reaction}. However, in the high time limit, the first decay of the survival probability is arising a nonintuitive nature of the decay probability. We are seeking to find a sufficient explanation of the strange behavior. Understandably, the slow decay of the probability with the increasing steepness of the potential is due to the domination of potential induced motion of the particle over the diffusion induced motion.

\section{CONCLUSION REMARKS}
The central result of this work is the semi-analytic expression for survival probability in Eq. (26). We are able to find the expression under satisfactory condition explicitly in one dimension, and it is confirming the nonmonotonic effect as it was perceived from the numerical analysis. Since our calculation is limited by the localized sink with infinite strength on the parabolic potential, placed at the uphill location relative to the initial distribution, so a general improvement over our derivation would be the solution for a sink of finite strength with the delocalized initial distribution for an active barrier process. The present result on the survival probability for the active barrier process would be the first-ever approach towards the analytic solution. Developing analytic model in this field is needful for the obvious reason that there are many processes involving diffusion in an active potential barrier in physics, chemistry and biology \citep{marcus1985electron,bagchi2012molecular}.       
\section{ACKNOWLEDGEMENT}
The author (C.S.) would like to thank the Indian Institute of Technology Mandi for  Half-Time  Research Assistantship fellowship. 
\section{APPENDIX: COMPARISON WITH THE EXACT FORM OF THE PROPAGATOR }
In the following, we provide a comparison of the time domain propagator of the Smoluchowski equation for different points with our approximated one. With considering the propagator $G(x,t|x_{0})$ with $x_{0}=0$ in Eq. (13), we can rewrite it in Eq. (14) employing the following approximation that
\begin{equation}
\begin{split}
     (1-Exp[-2 \gamma t])^{-1}\approx 1+Exp[-2 \gamma t], \quad&  t \neq 0\quad \text{and / or} \\
     & \gamma \neq 0. 
\end{split}
\end{equation}

The limiting value of these parameters in the above equation limits the study near to initial time or in case flat potential.
We can write the propagator in Eq. (13) for the other propagator, $\mathscr{G}_{0}(x_{1},s|x_{1})$ in Laplace domain as
\begin{equation}
    G(x_{1},t|x_{1})=  \sqrt{\frac{\gamma}{2 \pi D }}(1-Exp[-2 \gamma t])^{-1/2} Exp[-\frac{x_{1}^2 \gamma}{2 D}\times\frac{1-e^{-\gamma t}}{1+e^{-\gamma t}}].
\end{equation}
We see that the above equation may be modified into
\begin{equation}
    G(x_{1},t|x_{1})=  \sqrt{\frac{\gamma}{2 \pi D }}e^-\frac{x_{1}^2 \gamma}{2 D}(1-Exp[-2 \gamma t])^{-1/2},
\end{equation}
for the domination of $D$ over the steepness $\gamma$. This condition is also applicable for Eq. (14). A good coincidence of exact and approximated propagator is noticed, these are demonstrated in in Fig. 2 and in Fig.  3. The characterizations of the survival probability corresponding to the central result Eq. (23), are made under the same best fitting parameter as in the comparison of propagators. 




\refname


%% 
%% Copyright 2007-2019 Elsevier Ltd
%% 
%% This file is part of the 'Elsarticle Bundle'.
%% ---------------------------------------------
%% 
%% It may be distributed under the conditions of the LaTeX Project Public
%% License, either version 1.2 of this license or (at your option) any
%% later version.  The latest version of this license is in
%%    http://www.latex-project.org/lppl.txt
%% and version 1.2 or later is part of all distributions of LaTeX
%% version 1999/12/01 or later.
%% 
%% The list of all files belonging to the 'Elsarticle Bundle' is
%% given in the file `manifest.txt'.
%% 
%% Template article for Elsevier's document class `elsarticle'
%% with harvard style bibliographic references

%%\documentclass[preprint,12pt,authoryear]{elsarticle}
\documentclass[review]{elsarticle}
%% Use the option review to obtain double line spacing
%% \documentclass[authoryear,preprint,review,12pt]{elsarticle}

%% Use the options 1p,twocolumn; 3p; 3p,twocolumn; 5p; or 5p,twocolumn
%% for a journal layout:
%% \documentclass[final,1p,times,authoryear]{elsarticle}
%% \documentclass[final,1p,times,twocolumn,authoryear]{elsarticle}
%% \documentclass[final,3p,times,authoryear]{elsarticle}
%% \documentclass[final,3p,times,twocolumn,authoryear]{elsarticle}
%% \documentclass[final,5p,times,authoryear]{elsarticle}
%% \documentclass[final,5p,times,twocolumn,authoryear]{elsarticle}

%% For including figures, graphicx.sty has been loaded in
%% elsarticle.cls. If you prefer to use the old commands
%% please give \usepackage{epsfig}

%% The amssymb package provides various useful mathematical symbols
\usepackage{amssymb}
\usepackage[utf8]{inputenc}
\usepackage{natbib}
\usepackage{graphicx}
\usepackage{caption}
\usepackage{subcaption}
\usepackage{tabularx}
\usepackage{color}
\usepackage{cite}
\usepackage{lineno}
\usepackage{hyperref}
\usepackage{multirow}
\usepackage[final]{pdfpages}
\usepackage{fancyvrb}
\usepackage{amsmath,amsfonts,amsthm,amssymb}
\usepackage{setspace}
\usepackage{Tabbing}
\usepackage{color}
\usepackage{fancyhdr}
\usepackage{lastpage}
\usepackage{extramarks}
\usepackage{chngpage}
\usepackage{soul,color}
\usepackage{graphicx,float,wrapfig}
\usepackage{framed}
\usepackage{color}
\usepackage{epstopdf}
\usepackage{capt-of}
\usepackage{mathrsfs} 
\usepackage{graphicx}
\usepackage{braket}
\setcitestyle{square}
%%\usepackage{biblatex}
%% The amsthm package provides extended theorem environments
%% \usepackage{amsthm}

%% The lineno packages adds line numbers. Start line numbering with
%% \begin{linenumbers}, end it with \end{linenumbers}. Or switch it on
%% for the whole article with \linenumbers.
%% \usepackage{lineno}

\journal{Physica A}

\begin{document}

\begin{frontmatter}


 \author{Chinmoy Samanta}
 \ead{samantachinmoy111@gmail.com}

\title{Reaction-Diffusion dynamics in presence of  active barrier: Pinhole sink}

\address{School of Basic Sciences, Indian Institute of Technology Mandi,
Kamand, Himachal Pradesh, 175005, India }

\begin{abstract}
\noindent In this article, we give a semi-analytic expression for survival probability when particles are diffusing in an active potential well. There is no analytic solution available in the literature, due to the requirement of inverse Laplace transform of the propagator, when a sink is placed at the uphill of the parabolic potential even in case of the localized sink. We also explain some of the physical aspects by using our solution. 
\end{abstract}

\begin{keyword}


Statistical physics, Smoluchowski equation, Active barrier, Semi-Analytical model.
\end{keyword}

\end{frontmatter}


\section{INTRODUCTION}
For a long time, the study of diffusion of a particle in the potential well facing active barrier has been an interesting aspect in reaction-diffusion theory \citep{burada2012escape,bagchi2012molecular,hanggi1990reaction,kramers1940brownian,skinner1978relaxation,grote1980stable}. Even the one dimensional model itself is employed to various chemical and biological phenomena involving chemical reaction in condensed phase, electron \citep{marcus1985electron}, and proton \citep{samanta1990deuterium} transfer processes. In semiconductor physics, electron and hole are facing active barrier for the conduction of current \citep{tamura2013potential}, tunneling in semiconductor \citep{chang1974resonant}. From the theoretical point of view, the problem is to calculate the probability that the particle will still be in the excited state, after a time t. The major obstacle of developing the theory for similar kind of phenomena was the complex form of the propagator of the relevant equation, and that hinder the finding of the Laplace transform even in consideration of localized initial distribution as well as localized sink. In ref. \citep{samanta1993exact}, they provided the equation in the Laplace domain for calculating the rate constant. However, the exact expression for survival probability is still not unveiled. Although, recently, a real-time numerical result of the probability is provided by Spendier $et$ $al.$ \citep{spendier2013reaction}. In Fig 4, they showed that the character of the survival probability is deviating drastically for the uphill and downhill location of the sink with respect to the initial particle location. The fascinating nonmonotonic effect has been noticed when particle faces active barrier (uphill placement of sink) with varying steepness of the parabolic potential.
\newline In this article we provide a semi-analytic expression for the survival probability $Q(t)$ of the particle undergoing in parabolic potential and the localized sink is placed at the uphill relative to centrally placed initial distribution. In order to get the expression in the time domain, we have considered an approximation of the Smoluchowski time domain propagator but remain the same in the long-time limit. So our study is limited by beyond the initial time. An interesting nonmonotonic phenomenon is observed as in the ref. \citep{spendier2013reaction}. The exact and approximate form of the propagator is compared graphically in the Appendix.  

\section{SMOLUCHOWSKI EQUATION AND SURVIVAL PROBABILITY}
We begin with a generalization of the diffusion equation, viz., the Smoluchowski equation,
\begin{equation}
    \frac{\partial P(x,t)}{\partial t} = \frac{\partial}{\partial x}(\gamma x P(x,t)+D \frac{\partial P(x,t)}{\partial x})-S(x)P(x,t).
\end{equation}
In the above $P(x,t)$ describe the probability density of finding the particle, executing random walk, at position $x$ and time $t$. $D$ stands for diffusion coefficient, causes diffusive motion. $S(x)$ is position dependent sink term on the potential energy curve (PEC). The rate of motion towards the centre of attractive potential is defined by $\gamma$ and the motion can be called as potential induced motion. The propagator of the above equation in the absence of sink term is nothing but Green's function of the equation can be found by the method of characteristics[] with considering the initial feeding is localized, $P(x,0)=\delta(x-x_{0})$, which has the form 
\begin{equation}
    G(x,t|x_{0})=\frac{Exp[-(x-x_{0} Exp[-\gamma t])^{2}/4D\tau(t)]}{\sqrt{4 \pi D \tau(t)}},
\end{equation}
where $\tau$ is given by
\begin{equation}
    \tau(t)=\frac{1-Exp[-2 \gamma t]}{2 \gamma}.
\end{equation}
The survival probability is 
\begin{equation}
    Q(t)=\int_{-\infty}^{+\infty}P(x,t)dx.
\end{equation}
 The Laplace transform of Eq. (1), which is  
\begin{equation}
     [s-\mathscr{L}+S(x)]\mathscr{P}(x,s)=P(x,0),
\end{equation}
where
\begin{equation}
      \mathscr{L}=(D\frac{\partial^2}{\partial x^2}+\frac{\partial}{\partial x}\gamma x).
\end{equation}
Here $\mathscr{P}(x,s)$ is the Laplace transform of $P(x,t)$. The solution in terms of Green's function of the Eq. (5) is given by
\begin{equation}
     \mathscr{G}(x,s|x_{0})=<x| [s-\mathscr{L}+S(x)+]^{-1}|x_{0}>.
     \end{equation}
By using the operator identity of quantum mechanics we can get
     \begin{equation}
         [s-\mathscr{L}+S(x)]^{-1}=[s-\mathscr{L}]^{-1}-[s-\mathscr{L}]^{-1}S(x)[s-\mathscr{L}+S(x)]^{-1}.
         \end{equation}
For a localized capture of the particle the sink function is chosen to be Dirac delta function, $S(x)= k_{1}\delta(x-x_{1})$. $k_{1}$ is being the strength of the sink function. Thus,       
\begin{equation}
\mathscr{G}(x,s|x_{0})=\mathscr{G}_{0}(x,s|x_{0})-\frac{k_{1}\mathscr{G}_{0}(x,s|x_{1})\mathscr{G}_{0}(x_{1},s|x_{0})}{1+k_{1}\mathscr{G}_{0}(x_{1},s|x_{1})}.
\end{equation}
In the above $\mathscr{G}_{0}(x,s|x_{0})$, the Green's function corresponding to no sink on the PEC and it is nothing but Laplace transform of $G(x,t|x_{0})$, is defined as\\
\begin{equation}
    \mathscr{G}_{0}(x,s|x_{0})=<x|[s-\mathscr{L}]^{-1}|x_{0}>.
\end{equation}
The survival probability in the Laplace-domain would be
\begin{equation}
    \tilde{Q}(s)=\frac{1}{s}[1-\frac{\mathscr{G}_{0}(x_{1},s|x_{0})}{\frac{1}{k_{1}}+\mathscr{G}_{0}(x_{1},s|x_{1})}].
\end{equation}
Our calculation start with combining Eq. (11) and Eq. (2) In the next section.

\section{NON CENTRALLY PLACED TRAP: APPROX. ANALYTIC SOLUTION}
Due to the complex form of the Eq. (2), it is difficult to get the analytical solution in the time domain even for the localized sink of arbitrary position. Although methods are available to get the analytic solution in Laplace domain for the generalized sink (by considering collection Dirac delta function) on PEC. We consider the case when the initial distribution is localized at the center of the potential(consider to be the origin without loss of generality), and position of the localized sink is anywhere except the center. Therefore Eq. (11) become
\begin{equation}
    \tilde{Q}(s)=\frac{1}{s}[1-\frac{\mathscr{G}_{0}(x_{1},s|0)}{\frac{1}{k_{1}}+\mathscr{G}_{0}(x_{1},s|x_{1})}].
\end{equation}
We need to find the Laplace transform of $G(x,t|x_{0})$ by putting $x_{0}=0$ in order to get $\mathscr{G}_{0}(x_{1},s|0)$. We have the expression 
\begin{equation}
    G(x,t|0)= \sqrt{\gamma/(2 \pi D)} (1-Exp[-2 \gamma t])^{-1/2} Exp[-(x^{2}\gamma/(2 D))/(1-Exp[-2 \gamma t])].
\end{equation}
In case of the limit beyond initial time, we can write 


\begin{equation}
    G(x_{1},t|0)=  \sqrt{\frac{\gamma}{2 \pi D }}e^{-\frac{x_{1}^2 \gamma}{2 D}}(1-Exp[-2 \gamma t])^{-1/2} Exp[-\frac{x_{1}^2 \gamma}{2 D}\times Exp[-2 \gamma t]].
\end{equation}

One can find the Laplace transform of Smoluchowski propagator, the above equation, from a table of Laplace transform \citep{bateman1954tables,roberts1966table} or otherwise, that
\begin{equation}
    \mathscr{G}_{0}(x_{1},s|0)= \frac{e^{-\frac{3 x_{1}^{2} \gamma}{4 D}}}{\sqrt{8 \pi D \gamma}} \frac{\Gamma(\frac{1}{2}) \Gamma (\frac{s}{2\gamma})}{\Gamma(\frac{1}{2}+ \frac{s}{2 \gamma})}(-\frac{x_{1}^{2}\gamma}{2 D})^{-\frac{1}{4}(1+\frac{s}{\gamma})}M_{1/4-s/4\gamma,s/4 \gamma-1/4}(-\frac{x_{1}^{2}\gamma}{2 D}).
\end{equation}
In the above $\Gamma(n)$ is the gamma function and the $M$ in Eq. (16) is the Whittaker $M$ function define in ref. \citep{abramowitz1965handbook} as 
\begin{equation}
    M_{\kappa,\mu}(z)=e^{-z/2}z^{\mu+1/2}M(1/2+\mu-\kappa,1+2\mu,z),
\end{equation}
where
\begin{equation}
    M(a,b,z)=\frac{\Gamma(b)}{\Gamma(b-a)\Gamma(a)}\int_{0}^{1}dt e^{z t}t^{a-1}(1-t)^{b-a-1}.
\end{equation}
For the case of other propagator, $\mathscr{G}_{0}(x_{1},s|x_{1})$, of Eq. (12) under the approximation, it becomes 
\begin{equation}
\mathscr{G}_{0}(x_{1},s|x_{1})=\frac{e^{-\frac{x^2 \gamma}{2 D}}}{ \sqrt{8 \pi D \gamma}} B(\frac{s}{2 \gamma},\frac{1}{2}),
\end{equation}
where $B(z,w)=\Gamma(z)\Gamma(w)/\Gamma(z+w)$ is the beta function. We consider the case of perfect absorption for that $k_{1}\rightarrow{}\infty$. Substitution of Eq. (20) and Eq. (16) into Eq. (12) gives an expression for survival probability in Laplace domain, which is
\begin{equation}
    \tilde{Q}(s)=\frac{1}{s}-\frac{e^{-\frac{x^2 \gamma}{4 D}}\frac{\Gamma(\frac{1}{2}) \Gamma (\frac{s}{2\gamma})}{\Gamma(\frac{1}{2}+ \frac{s}{2 \gamma})}(-\frac{x_{1}^{2}\gamma}{2 D})^{-\frac{1}{4}(1+\frac{s}{\gamma})}M_{1/4-s/4\gamma,s/4 \gamma-1/4}(-\frac{x_{1}^{2}\gamma}{2 D})}{s B(\frac{s}{2 \gamma},\frac{1}{2})}.
\end{equation}
\begin{figure}[hbt!]
    \centering
    \includegraphics[width=8cm]{active_barrier.jpg}
   \caption{\small{\emph{Survival probability $Q(t)$ Vs $t$ with $x1 = 2;
D = 100$. We see the interesting feature when the placement of initial distribution is at the center of the potential and the trap is elsewhere. We get the nonmonotonic effect as the decay of $Q(t)$ is decreased by increasing the potential steepness but then enhanced on further increase in time. }}}
\end{figure}
The denominator of the second term can be written as
\begin{equation}
    \frac{1}{s B(\frac{s}{2 \gamma},\frac{1}{2})} = \frac{1}{2 \gamma ( \Gamma(1/2))^2}B(\frac{s}{2 \gamma}+\frac{1}{2},\frac{1}{2}).
\end{equation}
Farther we can write the Eq. (21) as
\begin{equation}
     \tilde{Q}(s)=\frac{1}{s}-\tilde{f}_{1}(s)\tilde{f}_{2}(s)
\end{equation}
where
\begin{equation}
\begin{split}
     \tilde{f}_{1}(s)&=\frac{\Gamma(\frac{1}{2}) \Gamma (\frac{s}{2\gamma})}{\Gamma(\frac{1}{2}+ \frac{s}{2 \gamma})}(-\frac{x_{1}^{2}\gamma}{2 D})^{-\frac{1}{4}(1+\frac{s}{\gamma})}e^{-\frac{x^2 \gamma}{4 D}}M_{1/4-s/4\gamma,s/4 \gamma-1/4}(-\frac{x_{1}^{2}\gamma}{2 D});\\
     \tilde{f}_{2}(s)& = \frac{B(\frac{s}{2 \gamma}+\frac{1}{2},\frac{1}{2})}{2 \gamma(\Gamma(1/2))^2}.
\end{split}
\end{equation}
After doing inverse Laplace transform from the table we get the general semi analytic expression in convolution form for $Q(t)$ is given by
\begin{equation}
    Q(t)=1- \int_{0}^{t}f_{1}(t_{0})f_{2}(t-t_{0})dt_{0}.
\end{equation}

Here $f_{1}(t)$ and $f_{2}(t)$ are inverse Laplace transform of $\tilde{f}_{1}(s)$ and $\tilde{f}_{2}(s)$ respectively. It is easy to find the inverse Laplace transform of $\tilde{f}_{1}(s)$ and we use a property of inverse Laplace transform to find $f_{2}(t)$ from $\tilde{f}_{2}(s)$ \citep{roberts1966table}. The property is 
\begin{equation}
    L^{-1}[\tilde{g}(as-b)]=\frac{1}{a}g(\frac{t}{a})e^{\frac{b}{a}t},
\end{equation}
where a, b are constant. The explicit form of $f_{1}(t)$ and $f_{2}(t)$ are
\begin{equation}
    \begin{split}
        f_{1}(t) & =\frac{2 \gamma}{\sqrt{1-Exp[-2 \gamma t]}}Exp[-\frac{x_{1}^{2}\gamma}{2 D}Exp[-2 \gamma t]];\\
         f_{2}(t) & =\frac{Exp[- \gamma t]}{(\Gamma(1/2))^2\sqrt{1-Exp[-2 \gamma t]}}
    \end{split}
\end{equation}
\begin{figure}[hbt!]
    \centering
    \includegraphics[width=8cm]{g10.jpg}
    \caption{\small{\emph{Actual (blue) and approx.(red) value of $G(x_{1},t|x_{0})$  Vs $t$ for different value of $D$. From the above to the bottom, the curves correspond to $D=10$, $50$ and $100$ respectively. Other fixed parameters are $\gamma = 1$, $x_{1} = 2$ and $x_{0} = 0$.}}}
\end{figure}
\begin{figure}[hbt!]
    \centering
    \includegraphics[width=8cm]{g11.jpg}
   \caption{\small{\emph{Actual (blue) and approx.(red) value of $G(x_{1},t|x_{1})$  Vs $t$ for different value of $D$. From the above to the bottom, the curves correspond to $D=10$, $50$ and $100$ respectively. Other fixed parameters are $\gamma = 1$ and $x_{1} = 2$.}}}
\end{figure}
The characteristics of survival probability with the different strength of the potential are demonstrated in Fig. 1. As it is prohibited to consider the short time limit of our analysis, we interpret in the high time limit. It is merely depicted that the behavior of $Q(t)$ is as expected as in the ref. \citep{spendier2013reaction}. However, in the high time limit, the first decay of the survival probability is arising a nonintuitive nature of the decay probability. We are seeking to find a sufficient explanation of the strange behavior. Understandably, the slow decay of the probability with the increasing steepness of the potential is due to the domination of potential induced motion of the particle over the diffusion induced motion.      



\section{CONCLUSION REMARKS}
The central result of this work is the semi-analytic expression for survival probability in Eq. (26). We are able to find the expression under satisfactory condition explicitly in one dimension, and it is confirming the nonmonotonic effect as it was perceived from the numerical analysis. Since our calculation is limited by the localized sink with infinite strength on the parabolic potential, placed at the uphill location relative to the initial distribution, so a general improvement over our derivation would be the solution for a sink of finite strength with the delocalized initial distribution for an active barrier process. The present result on the survival probability for the active barrier process would be the first-ever approach towards the analytic solution. Developing analytic model in this field is needful for the obvious reason that there are many processes involving diffusion in an active potential barrier in physics, chemistry and biology \citep{marcus1985electron,bagchi2012molecular}.       
\section{ACKNOWLEDGEMENT}
The author (C.S.) would like to thank the Indian Institute of Technology Mandi for  Half-Time  Research Assistantship fellowship. 
\section{APPENDIX: COMPARISON WITH THE EXACT FORM OF THE PROPAGATOR }
In the following, we provide a comparison of the time domain propagator of the Smoluchowski equation for different points with our approximated one. With considering the propagator $G(x,t|x_{0})$ with $x_{0}=0$ in Eq. (13), we can rewrite it in Eq. (14) employing the following approximation that
\begin{equation}
\begin{split}
     (1-Exp[-2 \gamma t])^{-1}\approx 1+Exp[-2 \gamma t], \quad&  t \neq 0\quad \text{and / or} \\
     & \gamma \neq 0. 
\end{split}
\end{equation}

The limiting value of these parameters in the above equation limits the study near to initial time or in case flat potential.
We can write the propagator in Eq. (13) for the other propagator, $\mathscr{G}_{0}(x_{1},s|x_{1})$ in Laplace domain as
\begin{equation}
    G(x_{1},t|x_{1})=  \sqrt{\frac{\gamma}{2 \pi D }}(1-Exp[-2 \gamma t])^{-1/2} Exp[-\frac{x_{1}^2 \gamma}{2 D}\times\frac{1-e^{-\gamma t}}{1+e^{-\gamma t}}].
\end{equation}
We see that the above equation may be modified into
\begin{equation}
    G(x_{1},t|x_{1})=  \sqrt{\frac{\gamma}{2 \pi D }}e^-\frac{x_{1}^2 \gamma}{2 D}(1-Exp[-2 \gamma t])^{-1/2},
\end{equation}
for the domination of $D$ over the steepness $\gamma$. This condition is also applicable for Eq. (14). A good coincidence of exact and approximated propagator is noticed, these are demonstrated in in Fig. 2 and in Fig.  3. The characterizations of the survival probability corresponding to the central result Eq. (23), are made under the same best fitting parameter as in the comparison of propagators. 




\refname
\begin{thebibliography}{00}
\bibitem{burada2012escape}
P. S. Burada and B. Lindner, Phys. Rev. E. 85(2012)032102.
\bibitem{bagchi2012molecular}
B. Bagchi, \emph{Molecular Relaxation in liquids}(Oxford University Press, New York,USA, 2012).
\bibitem{hanggi1990reaction}
P. H{\"a}nggi, P. Talkner and M. Borkovec, Rev. Mod. Phys. 62(1990)251.
\bibitem{kramers1940brownian}
H. A. Kramers, Physica 7(1940)284.
\bibitem{skinner1978relaxation}
J. L. Skinner and P. G. Wolynes, J. Chem. Phys. 69(1978)2143.
\bibitem{grote1980stable}
R. F. Grote and J. T. Hynes, J. Chem. Phys. 73(1980)2715.
\bibitem{marcus1985electron}
R. A. Marcus and N. Sutin, Biochim. Biophys. Acta 811(1985)265.
\bibitem{samanta1990deuterium}
A. Samanta, S. K. Ghosh, H. K. Sadhukhan, Chem. Phys. Lett. 168(1990)410.
\bibitem{tamura2013potential}
H. Tamura and I. Burghardt, J. Phys. Chem. C 117(2013)15020.
\bibitem{chang1974resonant}
L. L. Chang, L. Esaki and R. Tsu, Appl. Phys. Lett. 24(1974)593.
\bibitem{samanta1993exact}
A. Samanta, S. K. Ghosh, Phys. Rev. E. 47(1993)4568.
\bibitem{spendier2013reaction}
K. Spendier, S. Sugaya and V.M. Kenkre, Phys. Rev. E. 88(2013)062142. 
\bibitem{bateman1954tables}
H. Bateman and A. Erd{\'e}lyi, \emph{Tables of Integral Transforms. Vol. 1}(McGraw-Hill Book Company, New York/Toronto/London, 1954).
\bibitem{roberts1966table}
 G. E. Roberts and H. Kaufman, \emph{Table of Laplace Transforms}
(W. B. Saunders Company, Philadelphia/London, 1966).
\bibitem{abramowitz1965handbook}
 M. Abramowitz and I. A. Stegun, \emph{Handbook of Mathematical
Functions} (Dover Publications, Toronto, 1970).

\end{thebibliography}
\end{document}
\endinput
%%
%% End of file `elsarticle-template-harv.tex'.
\end{document}